\documentstyle[tighten,pre,aps,epsf,amsmath]{revtex}

\begin{document}
\draft

\newcommand{\Gn}
    {\setlength{\fboxsep}{-.4pt}\fbox{\rule{1.5em}{0mm}\rule{0mm}{1ex}}}
\newcommand{\notGn}{\rule{1.5em}{1ex}}
\newcommand{\Prob}{{\rm Prob}}
\newcommand{\annil}{$A+A\to0$}
\newcommand{\Ai}{{\rm Ai}}
\newcommand{\Bi}{{\rm Bi}}
\newcommand{\cs}{c_{\rm s}}
\newcommand{\Gs}{G_{\rm s}}

\title{A Method of Intervals for the Study of Diffusion-Limited 
Annihilation, \annil}

\author{Thomas O. Masser and Daniel ben-Avraham\footnote{{\bf e-mail:}
benavraham@clarkson.edu}}
\address{Physics Department and Clarkson Institute for Statistical
Physics (CISP), Clarkson University, Potsdam, NY 13699-5820}

\maketitle

\begin{abstract} 
We introduce a method of intervals for the analysis of diffusion-limited
annihilation, $A+A\to 0$, on the line.  The method leads to manageable
diffusion equations whose interpretation is intuitively clear.  As an
example, we treat the following cases: (a)~annihilation in the infinite
line and in infinite (discrete) chains; (b)~annihilation with input of
single particles, adjacent particle pairs, and particle pairs separated
by a given distance; (c)~annihilation, $A+A\to0$, along with the birth
reaction $A\to3A$, on finite rings, with and without diffusion.

\end{abstract}

% INTRODUCTION
\section{Introduction}
\label{sec:intro}
Diffusion-limited reactions display a wide range of behavior characteristic
of nonequilibrium dynamics, such as self-organized criticality,
pattern formation, and dynamic phase transitions.  However, the exact
description of this behavior can only be determined for the simplest
reaction schemes.  Single-species annihilation, $A+A\to 0$, and
coagulation, $A+A\to A$, are among the few examples of
exactly solvable diffusion-limited reactions.  The Method of Empty
Intervals has provided many exact results concerning diffusion-limited
coagulation~\cite{dba90,doering91,doering92,dba95,dba98}.  Its advantage 
lies in the fact that it leads to simple diffusion equations which are
easy to solve, and whose interpretation is intuitively obvious.  We
develop a similar method of intervals for the study of diffusion-limited
annihilation reactions. 

Diffusion-limited annihilation has been studied in a variety of forms.
For the basic annihilation process, the exact time-dependent
concentration of particles has been determined for various initial
distributions on the infinite one-dimensional lattice and the continuous
real
line~\cite{torney83,balding88,spouge88,henkel,simon,droz,%
bramson,peliti,krapivsky93,privman,glaser}. 
The inclusion of particle input in the reaction scheme allows for a
nontrivial steady state concentration.  Both input of adjacent particle
pairs~\cite{racz85,lushnikov87} and single particle
input~\cite{racz85,martin95,rey97} have been studied.  A nontrivial
steady state may also occur when a birth reaction is included.
Annihilation with the symmetric birth reaction $A\rightarrow 3A$ has been
studied for its implications to interacting particle systems and
universality
theories~\cite{grassberger84,grassberger89,sudbury90,takayasu92,jensen93,%
dba94,mussawisade98,szabo99}.  Sudbury~\cite{sudbury90} referred to this
reaction as the Double Branching Annihilating Random Walk (DBARW). 
Without diffusion, the process is known as the Double Branching
Annihilating Process (DBAP).  We show how a
method of intervals, previously introduced in the
literature~\cite{takayasu92,alemany95,dba00,masser00,henkel01}, may be
used to model all of the above reactions. This powerful yet simple method
yields insights that extend beyond the known results~\cite{habib}.

The rest of this article is organized as follows.  In
section~\ref{sec:method} we introduce the method of intervals, as adapted
for the annihilation process \annil, and reproduce its well-known
kinetics.  Section~\ref{sec:input} deals with various cases of input.
Input of particle pairs separated by a distance
$y$ is  considered, and the limiting cases of input
of single particles ($y\to\infty$) and adjacent particle pairs ($y\to0$)
are examined. We also discuss input at a rate proportional
to a function of the global concentration of particles.  In
section~\ref{sec:birth} we present our results for DBARW and DBAP.  We
conclude with a discussion and open directions of research suggested by
the new method of intervals, in section~\ref{sec:discussion}.

% The Gn method
\section{The method of odd/even intervals}
\label{sec:method}
We now adapt the method of empty intervals, conceived
originally for the analysis of diffusion-limited coalescence, $A+A\to A$,
to reactions such as annihilation, \annil, where the parity (the number of
particles modulo $2$) is conserved.  Consider the process
\annil, taking place in a one-dimensional lattice. 
The particles hop randomly to the nearest site on their right or left, at
equal rate $\Gamma$, and annihilate immediately upon encounter.  Let
$G_n(t)$ be the probability that an arbitrary segment of $n$ consecutive
sites contains an even number of particles, at time $t$.  (We assume, for
the moment, that the system is infinite and homogeneous.)  A site can be
either empty or occupied by a single particle, so the probability that a
site is occupied, {\em i.e.}, the particle density, is
\begin{equation}
\rho(t)=1-G_1(t)\;.  
\end{equation}

Since the reaction \annil\ conserves parity, the only way that $G_n$
might change is when particles at the edge of the segment hop outside,
or when particles just outside of the segment hop inside.  To describe
these events, we require
$F_n(t)$ --- the probability that an $n$-segment containing an even
number of particles is followed by the presence of a particle at the
$(n+1)$-th site.  This can be expressed in terms of the $G_n$ (Figure~1):
\begin{equation} 
\label{Fn}
2F_n(t)=(1-G_1)+(G_n-G_{n+1}) \;.     
\end{equation}
Likewise, $H_n(t)$ --- the probability that an $n$-segment containing an
{\it odd\/} number of particles is followed by a particle at the
$(n+1)$-th site --- is
\begin{equation}
\label{Hn}
2H_n(t)=(1-G_1)-(G_n-G_{n+1}) \;.     
\end{equation}

The evolution equation for $G_n$ is now readily obtained:
\begin{equation}
\label{Gn.Eq.FnHn}
{\partial\over\partial t}G_n(t)=
2\Gamma(F_{n-1}-H_{n-1}+H_n-F_n)\;.  
\end{equation}
The term proportional to $F_{n-1}$ on the r.h.s.\ expresses the event that
a particle at site $n$ jumps out of the segment, leaving an even number of
particles in the remaining $n-1$ sites (and hence in the $n$-segment);
$H_{n-1}$ corresponds to the same case, but when there are initially an
even number of particles in the $n$-segment (that is, an odd number in the
$(n-1)$-segment); $H_n$ and $F_n$ pertain to a particle just
outside of the $n$-segment, at site
$n+1$, jumping in.  The factor of
$2$ accounts for events taking place at {\em both}
edges of the segment, at equal rate.  Using Eqs.~(\ref{Fn}) and
(\ref{Hn}), this becomes
\begin{equation}
\label{Gn.eq}
{\partial\over\partial t}G_n(t)=
2\Gamma(G_{n-1}-2G_n+G_{n+1})\;.  
\end{equation}
The case of $n=1$ requires a special equation, since $G_0$ is undefined. 
Taking into account all the ways $G_1$ might change, one finds 
\begin{equation}
\label{G1.eq}
{\partial\over\partial t}G_1(t)=
2\Gamma(1-2G_1+G_2)\;.   
\end{equation}
Thus, Eq.~(\ref{Gn.eq}) may be understood to be valid for all $n\geq
1$, provided that one uses the boundary condition
\begin{equation}
\label{G_0.bc}
G_0(t)=1\;.  
\end{equation}
Additionally, since the $G_n$ are {\it probabilities\/}, we have
\begin{equation}
\label{Gn.bc} 
0\leq G_n(t)\leq 1\;. 
\end{equation}

The initial distribution of particles determines the required initial
condition, $G_n(0)$.  For example, suppose that at the start of the
process each site is occupied randomly and independently of other sites,
with probability $\rho_0$.  Then
\begin{equation}
\label{Gn.ic}
G_n(0)=\frac{1}{2}+\frac{1}{2}(1-2\rho_0)^n\;.
\end{equation}

Eq.~(\ref{Gn.eq}) may be solved by standard techniques, for
example, by Laplace-transforming with respect to time, fitting an 
exponential solution to the resulting difference equation, and finally
inverting the Laplace-transformed solution~\cite{abad99}.  With the
boundary conditions~(\ref{G_0.bc}) and (\ref{Gn.bc}), and the natural
initial condition~(\ref{Gn.ic}), one obtains:
\begin{equation}
\label{Gn.sol}
G_n(\tau)=\frac{1}{2}+\frac{1}{2}(1-2\rho_0)^ne^{\beta\tau}+
  \frac{n}{2}\int_0^{\tau}e^{-2\tau'}
  (1-e^{\beta\tau-\beta\tau'})I_n(2\tau')
  \,\frac{d\tau'}{\tau'}\;,
\end{equation}
where $\tau=2\Gamma t$, $\beta=4\rho_0^2/(1-2\rho_0)$, and $I_n(\cdot)$ is
the modified Bessel function of order $n$~\cite{AS}.  In particular, the
probability that a site contains a particle, is
\begin{equation}
\label{rho.sol}
\rho(\tau)=1-G_1(\tau)=\frac{1}{2}-\frac{1}{2}(1-2\rho_0)e^{\beta\tau}-
  \frac{1}{2}\int_0^{\tau}e^{-2\tau'}
  (1-e^{\beta\tau-\beta\tau'})I_1(2\tau')
  \,\frac{d\tau'}{\tau'}\;. 
\end{equation}

It is often more convenient to assume a low initial density of
particles, and work in the continuum limit.  Let $a$ be the lattice
spacing.  Then, putting 
$x=na$, $G_n(t)\to G(x,t)$, and $\Gamma=D/a^2$ in Eq.~(\ref{Gn.eq}),
and taking the limit $a\to0$, one obtains the diffusion equation
\begin{equation}
\label{Gx.eq}
{\partial\over\partial t}G(x,t)=2D{\partial^2\over\partial x^2}G(x,t)\;,
\end{equation}
with the boundary conditions
\begin{mathletters}
\label{bcs}
\begin{eqnarray}
&& G(x,0)=1\;, \label{Gx0.bc} \\
&& 0\leq G(x,t)\leq 1\;. \label{Gx.bc}
\end{eqnarray}
\end{mathletters}
The particle concentration is obtained from 
\begin{equation}
\label{c.def}
c(t)=\lim_{a\to0}\frac{\rho(t)}{a}=    
      -\frac{\partial}{\partial x}G(x,t)|_{x=0}\;.
\end{equation}  
Consider, for example, an initial
concentration $c_0$ of randomly placed particles.  The initial condition
is obtained from~(\ref{Gn.ic}), putting $\rho_0=c_0a$ and passing to the
continuum limit:
\begin{equation}
\label{Gx.ic}
G(x,0)=\frac{1}{2}+\frac{1}{2}e^{-2c_0x}\;.
\end{equation}
Solving for the concentration is then straightforward:
\begin{equation}
\label{c.sol.simple}
c(t)=c_0e^{z^2}{\rm erfc}(z)\;,\qquad z=2c_0\sqrt{2Dt}\;.
\end{equation}
This is the known result, with its familiar long-time asymptotic behavior,
\begin{equation}
\label{c.asymp.simple}
c(t)\sim\frac{1}{\sqrt{8\pi Dt}}\;, \qquad t\to\infty\;.
\end{equation}

% Annihilation with input: (a) Single particle (b) pairs, (c) pairs - y
% 	   (d) input proportional to a function of c(t).
\section{Annihilation with input}
\label{sec:input}
Let us now include the possibility of input,
where empty sites become spontaneously occupied at a prescribed rate.  We
shall assume that the input of particles is homogeneous (translation
invariant).  Various cases of input may be
analyzed through the method of intervals within this restriction. 

\subsection{Input of single particles}
\label{sec:single}
Consider a random, homogeneous input of single
particles at constant rate
$R$ per unit space per unit time.   If the input site is already occupied
we assume that the particles react immediately and the
site becomes empty.  In other words, the state of
individual sites flips (from empty to occupied, and vise versa) at rate
$Ra$ per unit time.  This kind of input affects the rate
of change of $G_n$ thus:
\begin{equation}
\label{dGn.Ainput}
\Bigl(\frac{\partial G_n}{\partial t}\Bigr)_{\text{A-input}}=
   nRa(1-G_n)-nRaG_n= nRa(1-2G_n)\;.
\end{equation}  
The term $nRa(1-G_n)$ accounts for the increase in $G_n$ due to the input
of a particle to an $n$-site interval, initially containing an {\em odd}
number of particles, while $-nRaG_n$ is the (negative) change in $G_n$
when a particle is input into an initially {\em even} $n$-interval. 
Note that Eq.~(\ref{dGn.Ainput}) is valid for all $n\geq 1$, and so the
input does not affect the boundary condition~(\ref{Gn.bc}). In the
continuum limit, the change due to input is $\partial G/\partial
t=Rx(1-2G)$, so Eq.~(\ref{Gx.eq}) becomes
\begin{equation}
\label{G.single}
\frac{\partial}{\partial t}G(x,t)=
  2D\frac{\partial^2}{\partial x^2}G(x,t) + Rx(1-2G(x,t))\;,
\end{equation}
subject to the same boundary conditions as without input.  The
problem can be solved by substituting 
\[
K(x,t)=1-2G(x,t)\;.
\]
$K(x,t)$ then satisfies
\begin{eqnarray*}
&& \frac{\partial}{\partial t}K(x,t)
 = 2D\frac{\partial^2}{\partial x^2}K(x,t) - 2RxK(x,t) \;,\\
&&  K(0,t) = -1\;.
\end{eqnarray*}
Using the method of separation of variables, we write
\[
K(x,t)=\sum_{\lambda\geq0}a_{\lambda}K_{\lambda}(x)e^{-\lambda t}\;,
\]
where the eigenfunctions satisfy
\[
2D\frac{\partial^2}{\partial x^2}K_{\lambda}(x) = 
    (2Rx-\lambda)K_{\lambda}(x)\;.
\]
Thus,
\[
K_{\lambda}(x)=\Ai\left[ \Bigl(\frac{R}{D} \Bigr)^{1/3} x 
    - \frac{\lambda}{2(DR^2)^{1/3}} \right]\;,
\]
where $\Ai(\cdot)$ is the Airy function --- a solution to the equation 
$\Ai''(z)=z\Ai(z)$~\cite{AS}.  The boundary condition $K(0,t)=-1$
implies
$K_{0}(0)=-1$ and $K_{\lambda}(0)=0$ ($\lambda > 0$).  The
condition $K_{0}(0)=-1$ yields the steady state solution:
\begin{equation}
\label{Gs.single}
G_{s}(x)=
  \frac{1}{2}+\frac{\Ai\bigl[(\frac{R}{D})^{1/3}x\bigr]}{2\Ai(0)}\;,
\end{equation}
and the steady state concentration:
\begin{equation}
\label{cs.A}
\cs=-\frac{\partial}{\partial x}\Gs(x)|_{x=0}=
  \frac{|\Ai'(0)|}{2\Ai(0)}\Bigl(\frac{R}{D}\Bigr)^{1/3}\;.
\end{equation}
The condition $K_{\lambda}(0)=0$ imposes discrete eigenvalues: 
\[
\lambda_n=2(DR^2)^{1/3}|a_n|\;,
\]
where $a_n$ are the zeros of the Airy function.  The relaxation time
to the stationary state, $\tau$, is then given by the minimal eigenvalue:
\begin{equation}
\tau^{-1}=\lambda_{\rm min}=2(DR^2)^{1/3}|a_{1}|=4.6762(DR^2)^{1/3}\;.
\end{equation}

\subsection{Input of adjacent particle pairs}
\label{sec:adjacent}
Consider now homogeneous, random input of particle {\em pairs} to adjacent
sites: any two adjacent sites become occupied at rate $Ra=r$ per unit
time.  As before, if a target site is already occupied, the site becomes
empty as a result of input.   The situation is 
analogous to that of input of single particles, but the
parity of an interval is now only affected when input occurs at the
interval's edge --- when only one particle of the pair falls right inside
the interval. $G_n$ is increased by input at the
edge of an odd $n$-interval, and decreased by input at an
even interval (See Figure~2).  Thus, the rate of change of $G_n$ due
to pair input is
\begin{equation}
\Bigl(\frac{\partial}{\partial t}G_n\Bigr)_{\text{AA-input}}=
 2[ r(1-G_n)-rG_n ]=2r(1-2G_n)\;.
\end{equation}
In the continuum limit, $G(x,t)$ must satisfy
\begin{equation}
\frac{\partial}{\partial t}G(x,t)=
   2D\frac{\partial^2}{\partial x^2}G(x,t) + 2r(1-2G(x,t))\;,
\end{equation}
with the usual boundary condition, $G(0,t)=1$.  

Again, we use $K(x,t)=1-2G(x,t)$ to determine the solution.  $K(x,t)$
must satisfy 
\begin{eqnarray*}
&& \frac{\partial}{\partial t}K(x,t) 
     =   2D\frac{\partial^2}{\partial x^2}K(x,t) - 4rK(x,t)\;, \\
&& K(0,t)   =   -1\;.
\end{eqnarray*}
Expanding  $K(x,t)=\sum_{\lambda}K_{\lambda}(x)e^{-\lambda t}$, as
before, we find
\[
2D\frac{\partial^2}{\partial x^2}K_{\lambda}(x) = 
   (4r-\lambda)K_{\lambda}(x)\;,
\]
with $K_{0}(0)=-1$ and $K_{\lambda}(0)=0$ ($\lambda > 0$).
This leads to the stationary solution
\begin{equation}
\label{Gs.pair}
\Gs(x)=\frac{1}{2}+\frac{1}{2}e^{-\sqrt{\frac{2r}{D}}x}\;,
\end{equation}
and the stationary concentration
\begin{equation}
\label{cs.AA}
\cs=\frac{1}{2}\Bigl(\frac{2r}{D}\Bigr)^{1/2}.
\end{equation}
The relaxation spectrum is continuous: 
\begin{equation}
\lambda>4r\;;\qquad
K_{\lambda}(x)=\sin\Bigl(\sqrt{\frac{\lambda
-4r}{2D}}x\Bigr)\;,
\end{equation}
and the relaxation time is
\begin{equation}
\tau^{-1}=\lambda_{\rm min}=4r\;.
\end{equation}
These results, along with the single particle input results, are in
complete agreement with previous work by R\'acz~\cite{racz85} and
Lushnikov~\cite{lushnikov87}.

\subsection{Input of correlated pairs}
The method of intervals allows us to deal with more complicated input.  As
an example, we consider input of particle pairs, separated by $m$ lattice
spacings.  This input of {\em correlated} pairs 
interpolates between the two cases discussed so
far: for
$m=0$ the particle pairs are adjacent, just as in
section~\ref{sec:adjacent}, while the case of section~\ref{sec:single} is
recovered as $m\to\infty$, since then the
input particles cannot possibly affect each other and the correlation is
lost.

Suppose that the particle pairs are deposited at rate $Ra$ per unit
time (per site).  The rate of change of
$G_n$ due to input is now different for
$n\leq m$ and $n>m$:  
\begin{mathletters}
\label{input.AmA}
\begin{eqnarray}
&&\Bigl(\frac{\partial}{\partial t}G_n\Bigr)_{\text{A$\cdots$A-input}}
  =2Rna(1-2G_n)\;, \qquad\qquad\quad n\leq m\;,\\
&&\Bigl(\frac{\partial}{\partial t}G_n\Bigr)_{\text{A$\cdots$A-input}}
  =2R(m+1)a(1-2G_n)\;, \,\,\qquad n>m\;.
\end{eqnarray}
\end{mathletters}
Again, Eq.~(\ref{input.AmA}a) does not affect the usual boundary
condition $G_0=1$.  Adding these contributions to Eq.~(\ref{Gn.eq}) and
taking the continuum limit, we obtain
\begin{mathletters}
\label{input.AyA}
\begin{eqnarray}
&& \frac{\partial}{\partial t}G(x,t)=
  2D\frac{\partial^2}{\partial x^2}G(x,t)+2Rx(1-2G(x,t))\;, 
  \qquad x\leq y\;,\\
&& \frac{\partial}{\partial t}G(x,t)=
  2D\frac{\partial^2}{\partial x^2}G(x,t)+2Ry(1-2G(x,t))\;, 
  \qquad x > y\;,
\end{eqnarray}
\end{mathletters}
with the usual boundary condition $G(0,t)=1$.  Additionally, $G(x,t)$ and
$\partial G(x,t)/\partial x$ are continuous at $x=y$.

For simplicity, we analyze only the steady state.  We find,
\begin{mathletters}
\begin{eqnarray}
&& \Gs(x)=c_1\Ai(\alpha x)+c_2\Bi(\alpha x)+\frac{1}{2}\;,
  \qquad x\leq y\;,\\
&& \Gs(x)=c_3e^{-\sqrt{\alpha^3y}\,x}+\frac{1}{2}\;,
  \qquad\qquad\qquad\,\, x>y \;,
\end{eqnarray}
\end{mathletters}
where $\Bi(\cdot)$ is the independent, divergent solution to Airy's
equation~\cite{AS}, $\alpha=(2R/D)^{1/3}$, and
$c_1$, $c_2$, $c_3$, are constants obtained from the boundary conditions
at $x=0$ and $x=y$.  The steady state concentration is
\begin{equation}
\cs=\frac{|\Ai'(0)|}{2\Ai(0)}\Bigl(\frac{2R}{D}\Bigr)^{1/3}
    f\left[\Bigl(\frac{2R}{D}\Bigr)^{1/3}y\right]\;,
\end{equation}
where
\begin{equation}
\label{crossover}
f(z)=\frac{\sqrt{z}\,\Bi(z)+\Bi'(z)+\sqrt{3z}\,\Ai(z)+\sqrt{3}\,\Ai'(z)}
  {\sqrt{z}\,\Bi(z)+\Bi'(z)-\sqrt{3z}\,\Ai(z)-\sqrt{3}\,\Ai'(z)}
\to\begin{cases}
\frac{\sqrt{z}}{2}\;, & z\ll1\;,\\
1\;, & z\gg 1\;.
\end{cases}
\end{equation}
Eq.~(\ref{crossover}) expresses the crossover behavior between input of
single particles and adjacent particle pairs.  Indeed, when $y\to\infty$
($z\to\infty$), the particles are essentially uncorrelated, and the steady
state is the same as for single particles, of Eq.~(\ref{cs.A}), but with
$2R$ instead of $R$.  This is because the input of pairs introduces
particles at twice the rate of single particles.  The limit of $y\to0$
($z\to0$) yields
$\cs\sim\sqrt{(R/D)y}$, just as for adjacent particle pairs
(Eq.~\ref{cs.AA}).  The crossover between the two regimes occurs about
$(R/D)^{1/3}y\sim1$ (Figure~3).

\subsection{Input proportional to some global property}
We can also analyze input at a rate proportional to some global property
of the system.  Consider input of single
particles, as in section~\ref{sec:single}, but at a rate proportional to
a functional of $G(x,t)$: 
\[
R={\cal R}[G(x,t)]\;.
\]
Putting this rate into Eq.~(\ref{G.single}) yields a non-local
partial differential equation for $G(x,t)$, which is generally difficult
to solve.  However, the steady state may be obtained in the following
manner.  In the steady state limit, the rate $R_{\rm s}={\cal R}[\Gs(x)]$ 
is constant.  This constant can be found by solving
\begin{equation}
R_{\rm s}={\cal R}[\Gs(x|R_{\rm s})]\;,
\end{equation}
where  $\Gs(x|R)$ is the distribution given by
Eq.~(\ref{Gs.single}), assuming constant rate $R$.  Once the value of
$R_{\rm s}$ is known, the steady state distribution, $\Gs(x|R_{\rm
s})$, and steady state concentration, $\cs$, follow.  Notice that
the same procedure applies for other kinds of input, with minor changes:
for example, for input of adjacent particle pairs, $\Gs(x|R)$ is
given by Eq.~(\ref{Gs.pair}) rather than~(\ref{Gs.single}).

An amusing example is input of single particles at a rate proportional to
a power of the particle concentration:
\begin{equation}
R=R_0\Bigl(\frac{c(t)}{c(0)}\Bigr)^{\alpha}\;.
\end{equation}
In this case we can find $\cs$ directly, from Eq.~(\ref{cs.A}):
\[
\cs=\gamma\Bigl(\frac{R_{\rm s}}{D}\Bigr)^{1/3}\;,
  \qquad \gamma=\frac{|\Ai'(0)|}{2\Ai(0)}\;,
\]
or,
\begin{equation}
\label{alpha}
\cs=
\Bigl(\frac{\gamma^3}{c(0)^\alpha}\frac{R_0}{D}\Bigr)
  ^{1/(3-\alpha)}\;.
\end{equation}
The true meaning of Eq.~(\ref{alpha}), for different values of $\alpha$,
is revealed only from a careful analysis of the pertinent effective-rate
equation~\cite{dba90}:
\begin{equation}
\label{eff}
\frac{d}{dt}c=-k_1c^3+k_2c^{\alpha}\;,\qquad k_1\sim D\;,\quad 
  k_2\sim R_0/c(0)^{\alpha}\;.
\end{equation}
Eq.~(\ref{eff}) has a steady state similar to~(\ref{alpha}), but one can
clearly see that that state is stable only for $\alpha<3$.  For
$\alpha>3$ the solution is unstable --- instead, the system flows to one
of the stable states $c=0$ or $c=\infty$.  A particularly intriguing
case, in which fluctuations not reflected in the
mean-field analysis will surely play a central role, is that of
$\alpha=3$.

% A+A->0, A->3A (DBARW, DBAP)
\section{Annihilation with symmetric birth}
\label{sec:birth}
We now treat the process of annihilation, \annil, with the addition of the
back reaction $A\to3A$, where particles give birth to two new particles
at the adjacent sites to their left and right, at rate $\Omega$ (per
particle, per unit time).  If birth takes place onto a
site that is already occupied, annihilation is immediate and the site
becomes empty.  This process has been analyzed by
Sudbury~\cite{sudbury90}, who named it the double branching annihilating
random walk (DBARW).  Sudbury~\cite{sudbury90} also studied the double
branching annihilating process (DBAP), which is the same as the DBARW
except that the particles do not diffuse.  The DBARW is now commonly
referred to as the branching-annihilating walk with two offspring
($n=2$-BAW), but since we wish to distinguish between the processes with
and without diffusion we shall utilize Sudbury's nomenclature. 

Consider the DBARW taking place on an $N$-site
chain, with periodic boundary conditions (or an $N$-site ring).  Because
the ring is finite, $G_N$ is determined by the parity of the initial
number of particles, and remains constant throughout the process.  The
effect of diffusion on $G_n$ has already been discussed, and is
given by Eqs.~(\ref{Gn.eq}) and (\ref{G1.eq}): 
\[
\Bigl(\frac{\partial}{\partial t}G_n\Bigr)_{\text{Diff}}
  =2\Gamma(G_{n-1}-2G_n+G_{n+1})\;, \qquad\qquad\quad 1<n<N\;,
\]
and 
\[
\Bigl(\frac{\partial}{\partial t}G_1\Bigr)_{\text{Diff}}
  =2\Gamma(1-2G_1+G_2)\;, \qquad\qquad\quad n=1\;.
\]
Birth affects $G_n$ in a similar way to diffusion: the parity of
an $n$-segment changes only when a particle just inside or just outside
the segment gives birth.  Thus,
\[
\Bigl(\frac{\partial}{\partial t}G_n\Bigr)_{\text{Birth}}
  =2\Omega(G_{n-1}-2G_n+G_{n+1})\;, \qquad\qquad\quad 1<n<N-1\;.
\]
For $n=1$, birth into an empty site decreases $G_1$, while birth into
an occupied site increases $G_1$.  The two effects add up to
\[
\Bigl(\frac{\partial}{\partial t}G_1\Bigr)_{\text{Birth}}
  =2\Omega(G_2-G_1)\;, \qquad\qquad\quad n=1\;.
\]
Finally, for $n=N-1$, birth from a particle at the inner edge of the
$(N-1)$-segment is similar to the case of generic $n$, but birth from  a
particle outside of the segment is different: because the lattice is a
$N$-ring, a particle outside of the $(N-1)$-segment gives birth to {\em
two} particles inside it (one at each edge), and the parity does not
change.  Thus,
\[
\Bigl(\frac{\partial}{\partial t}G_{N-1}\Bigr)_{\text{Birth}}
  =2\Omega(G_{N-2}-G_{N-1})\;, \qquad\qquad\quad n=N-1\;.
\]
Putting the contributions from diffusion and from birth together, we get
\begin{mathletters}
\label{dbarw.eq}
\begin{eqnarray}
\frac{\partial}{\partial t}G_n&&=2(\Gamma+\Omega)(G_{n-1}-2G_n+G_{n+1})\;,
\qquad\qquad\quad\qquad\qquad\qquad\quad\,\,\, 1<n<N-1\;,\\
\frac{\partial}{\partial t}G_1&&=2\Gamma(1-2G_1+G_2)
  +2\Omega(G_2-G_1)\;,
\qquad\qquad\quad\qquad\qquad\qquad\,\, n=1\;,\\
\frac{\partial}{\partial t}G_{N-1}&&=2\Gamma(G_{N-2}-2G_{N-1}+G_N)
  +2\Omega(G_{N-2}-G_{N-1})\;,
\qquad\qquad\quad n=N-1\;,
\end{eqnarray}
\end{mathletters}

For simplicity, we focus on the steady state, where time derivatives
are zero.  (The transient behavior may be analyzed by 
methods similar to that of Section~\ref{sec:method}.)  The general
(steady state) solution of Eq.~(\ref{dbarw.eq}a) is
$G_n=An+B$, where $A$ and $B$ are constants.  Their explicit value is
determined from the boundary conditions~(\ref{dbarw.eq}b) and
(\ref{dbarw.eq}c):
\begin{mathletters}
\label{AB}
\begin{eqnarray}
\Omega A - \Gamma B &=& -\Gamma\;,\\
(N\Gamma+\Omega)A+\Gamma B &=& -\Gamma G_N\;.
\end{eqnarray}
\end{mathletters}
Thus,
\[
A=-\frac{\Gamma(1-G_N)}{N\Gamma+2\Omega}\;, \qquad
  B=1-\frac{\Omega(1-G_N)}{N\Gamma+2\Omega}\;,
\]
and the stationary particle density is
\begin{equation}
\label{rho_DBARW}
\rho_{\rm s}=1-G_1=1-A-B=\frac{(\Gamma+\Omega)(1-G_N)}{N\Gamma+2\Omega}\;.
\end{equation}
When the initial number of particles is even, $G_N=1$ and the system
gravitates into its absorbing, empty state --- a state from which it
cannot evolve any further.  If the initial number of particles is odd
($G_N=0$), the system can never reach the absorbing state.  However, in
this case the steady state density is barely larger than $1/N$ (for $N$
large): the system comes as close to extinction as possible.  This
generalizes the result of Sudbury that $\rho_{\rm s} = 0$ for
the DBARW in infinite lattices.  

Consider now the DBAP, which is the case of no diffusion, $\Gamma=0$.
In this case, Eqs.~(\ref{AB}) imply $A=0$, $B$ arbitrary, or
$G_n=\text{const}$.  Suppose that the initial number of particles is
infinite, so that $G_N$ is not defined (but, since $G_n$ is
constant, $0\leq G_N\leq 1$).  Then, taking the limit
$\Gamma\to0$ in~(\ref{rho_DBARW}), we find $0\leq\rho_{\rm s}\leq1/2$. 
This too agrees with Sudbury, who has shown that any homogeneous,
random initial distribution leads to $\rho_{\rm s}=1/2$, but other
distributions lead to steady states with $0\leq\rho_{\rm s}\leq1/2$.
According to Sudbury, the possible steady states are the borders of the
product measures $\nu_p$.  Indeed, the border of $\nu_p$ has constant
$G_n$:  the condition that there be an even number of borders in an
$n$-segment is equivalent to having the same states at the edges of the
segment in the original $\nu_p$ measure.  But the probability of this
event is $G_n=p^2+(1-p)^2$, independent of $n$.

An amusing case of the DBAP is when the initial number of particles is
finite.  In this case $G_N$ is known exactly.  Thus, if the initial
number of particles is odd ($G_N=0$), then the $\Gamma\to0$ limit of
Eq.~(\ref{rho_DBARW}) yields $\rho_{\rm s}=1/2$, in agreement with
Sudbury.  It is instructive to see how this steady state manifests itself
in the case of finite rings.  An example, of $N=6$, is shown in
Appendix~{\bf A}.  If the initial number of particles is even ($G_N=1$),
Eq.~(\ref{rho_DBARW}) suggests that
$\rho_{\rm s}=0$, even for finite
$N$.  This is however not true, and it can be shown~\cite{sudbury90,dba94}
that the number of particles remains bounded.  (For example, two adjacent
particles would propagate as a pair forever, diffusing without change.) 
In this case the limit $\Gamma\to0$ is singular: even the tiniest
amount of diffusion would land the system in the absorbing state, but
the system can never become empty {\em without} diffusion.

\section{Discussion} %%% DISCUSSION
\label{sec:discussion}
In conclusion, the method of intervals enables one to obtain
exact results for a large class of diffusion-limited annihilation models
in one-dimension. The usefulness of this approach stems from the direct
consideration of parity conservation in the annihilation process;
annihilation conserves the number of particles modulo~2.  Defining $G_n$
as the probability that
$n$ consecutive sites contain an even number of particles exploits this
constraint.  In some sense, this method of intervals is a generalization
of the method of empty intervals used for coagulation
processes~\cite{dba90,doering91,doering92,dba95,dba98}.  A further
generalization exists that simultaneously models the $q$-state Potts
model, coagulation and annihilation~\cite{masser00}.   

The results of this paper can be summarized as follows.  The well known
result $c(t)=c_{0}e^{8c_{0}^{2}Dt}\text{erfc}(\sqrt{8c_{0}^{2}Dt})$ for
diffusion-limited annihilation with an initial Poisson distribution of
particles in the real line was reproduced, and we have obtained an
analogous solution for the discrete case of a linear lattice.  We have
studied the nontrivial steady state concentration resulting from particle
input.  In particular, it was shown that a smooth crossover occurs in the
steady state concentration as the separation of input particle pairs
increases from zero to infinity; as particle pair input crosses over to
single particle input, the concentration dependence changes from
$(R/D)^{1/3}$ to
$(R/D)^{1/2}$.  Our final example involved annihilation with the
symmetric birth reaction
$A\to3A$.  The method of intervals allows one to fully examine the
kinetics of this reaction scheme in finite ringed lattices.  As an
example, we derived the exact nature of the steady state solution for even
and odd populations, with or without diffusion.

The method of odd/even intervals introduced here for the case
of annihilation could be extended along the same lines as the method of
empty intervals used for
coagulation~\cite{doering91,doering92,dba98,burschka,dba}. This will be
the subject of future work.  For example, the kinetic
phase transition associated with reversible coagulation,
$A+A\rightleftharpoons A$~\cite{burschka}, might also have an echo in
$A+A\to0$,
$A\to3A$, at least in the case of finite lattices, when there exists a
non-empty steady state. Nonhomogeneous systems could be
studied with the addition of one variable, by focusing on
$G_{nm}(t)$ --- the probability that the interval between sites $n$ and
$m$ contains an even number of particles~\cite{doering91}.  It might also
be possible to obtain multiple-point correlation functions by studying
the joint probability that several distinct intervals contain odd/even
numbers of particles~\cite{doering92,dba98}. 

A natural question to ask is whether the inter-particle distribution
function (IPDF) for annihilation can be computed by the odd/even interval
method, as is the case for coagulation (with the method of empty
intervals).  In the latter case, the IPDF --- the probability that the
empty space between two particles is of length $x$ --- is given by
$p(x,t)=c(t)^{-1}\partial^2 E(x,t)/\partial x^2$, where $E(x,t)$ is the
probability that an interval of length $x$ is empty~\cite{dba90}. 
Unfortunately, $\partial^2 G(x,t)/\partial x^2$ does not convey analogous
information.  We believe that the full hierarchy of multiple-point
correlation functions may be obtained through the odd/even interval
method, and since it provides a full description of the system, 
the IPDF could be obtained as well.  While this might work, in principle,
the actual computation seems impractical.  It would be desirable to
find a more straightforward way, based on the odd/even interval method, to
compute the IPDF.

\acknowledgments  
We thank the National Science Foundation for support, under grant 
PHY-9820569.

\appendix
\section{Steady state of DBAP in $(N=6)$-ring}
Consider the DBAP on a ring of $N=6$ sites, when there is initially an
odd number of particles.  The system may assume only one of $5$
configurations, $\sigma_i$ ($i=1,2,\dots,5$).  These configurations and
the transition rates between them are illustrated in Figure~4.
{}From the figure, we see that the probabilities $\pi_i$ of having state
$\sigma_i$, obey the rate equations
\begin{eqnarray*}
&&{\dot\pi}_1=-\pi_1+\pi_2\;,\\
&&{\dot\pi}_2=\pi_1-3\pi_2+\pi_3\;,\\
&&{\dot\pi}_3=2\pi_2-2\pi_3+2\pi_4\;,\\
&&{\dot\pi}_4=\pi_3-3\pi_4+3\pi_5\;,\\
&&{\dot\pi}_5=\pi_4-3\pi_5\;,
\end{eqnarray*}
where dots denote differentiation with respect to time.  These equations
are linearly dependent, because of the normalization condition
$\sum_i\pi_i=1$.  Supplementing the rate equations with this condition, we
find the steady state:
$\pi_1=3/16$, $\pi_2=3/16$, $\pi_3=6/16$, $\pi_4=3/16$, $\pi_5=1/16$. 
The $G_n$ may be computed for each configuration, by averaging over all
possible locations of the $n$-segment (Table~I).  We can now compute the
average $G_n$, by weighing the values in Table~I with the $\pi_i$ found
above:
$\langle G_n\rangle=\sum_i\pi_iG_n(\sigma_i)$.  This procedure yields
the expected result, $\langle G_n\rangle=1/2$ ($n=1,2,\dots,5$). 
Rings of other sizes may be analyzed in much the same way, though we
found no obvious generalization, beyond the simple fact that $\langle
G_n\rangle$ is always $1/2$.

\begin{figure}
\centerline{\epsfxsize=7cm \epsfbox{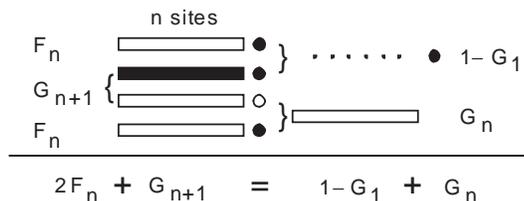}}
\caption{Computation of $F_n(t)$:
Empty/solid rectangles symbolize $n$-segments with an even/odd
number of particles. Empty/solid circles represent empty/occupied sites.}
\end{figure}

\begin{figure}
\centerline{\epsfxsize=7.5cm \epsfbox{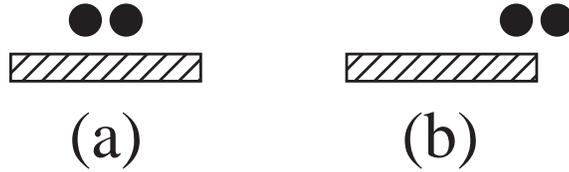}}
\caption{Change of $G_n$ with input of adjacent particle pairs: Hatched
rectangles represent $n$-segments with initially an even (odd) number of
particles.  (a)~Input of a pair inside the interval does not affect the
parity.  (b)~Input of the pair at the edge of the segment, when just one
particle lands inside, shifts the segment parity to odd (even).}
\end{figure}

\newpage
\begin{figure}
\centerline{\epsfxsize 12cm \epsfbox{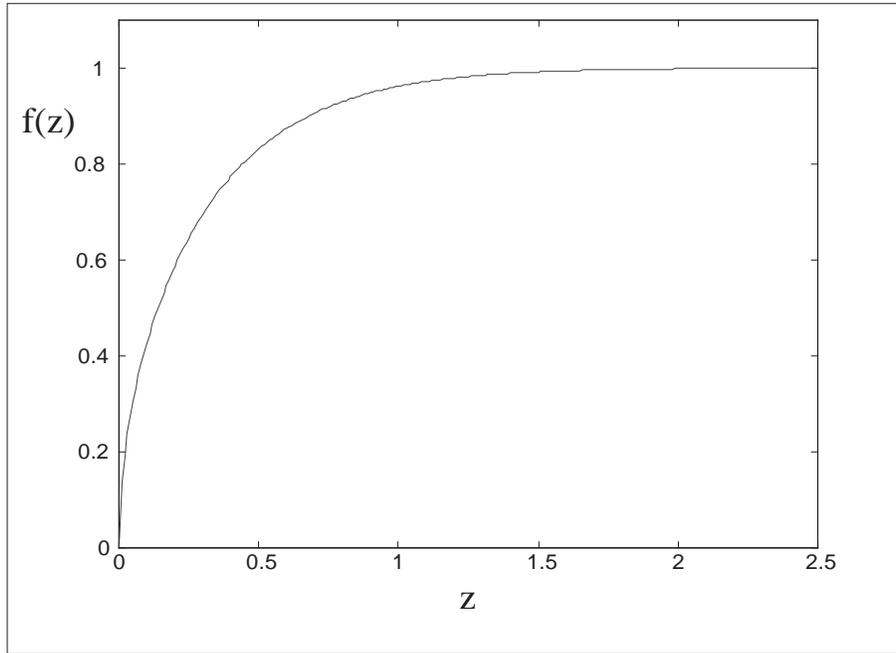}}
\caption{The crossover function $f(z)$ of Eq.~(\ref{crossover}).  Notice
the limiting behavior for small and large $z$ and the crossover about
$z\approx1$ evident from the plot.}
\end{figure}

\begin{figure}
\centerline{\epsfxsize=12cm \epsfbox{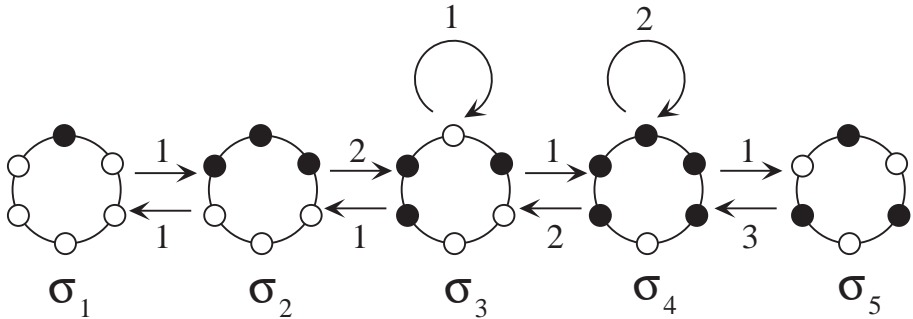}}
\caption{Configurations of DBAP in a $(N=6)$-ring: Empty/solid circles
symbolize empty/occupied sites.  The arrows and numbers indicate relative
transition rates between configurations.}
\end{figure}

\begin{table}
\narrowtext
\caption{$G_n$ for the configurations of DBAP in a $(N=6)$-ring, shown in
Figure~4}
\begin{tabular}{cccccc}
        &$\sigma_1$ &$\sigma_2$ &$\sigma_3$ &$\sigma_4$ &$\sigma_5$\\
\hline
$G_1$   &5/6        &3/6        &3/6        &1/6        &3/6        \\
$G_2$   &4/6        &4/6        &2/6        &4/6        &0          \\
$G_3$   &3/6        &3/6        &3/6        &3/6        &3/6        \\
$G_4$   &2/6        &2/6        &4/6        &2/6        &6/6        \\
$G_5$   &1/6        &3/6        &3/6        &5/6        &3/6        
\end{tabular}
\end{table}

\begin{thebibliography}{99} %%% REFERENCES


\bibitem{dba90} D. ben-Avraham, M. A. Burschka, and C.R. Doering,
``Statics and dynamics of diffusion-limited reactions: anomalous kinetics,
non-equilibrium self-ordering, and a dynamic transition," {\it J. Stat.
Phys.}  {\bf60}, 695 (1990).

\bibitem{doering91} C. R. Doering, M. A. Burschka, and W. Horsthemke,
``Fluctuations and correlations in a diffusion-reaction system: exact
hydrodynamics," {\it J. Stat. Phys.} {\bf 65}, 953 (1991).

\bibitem{doering92} C. R. Doering, ``Microscopic spatial correlations
induced by external noise in a reaction-diffusion system," {\it Physica A}
{\bf 188}, 386 (1992).

\bibitem{dba95} D. ben-Avraham, ``The Method of inter particle
distribution functions for diffusion-reaction systems in one-dimension,"
{\it Mod. Phys. Lett. B} {\bf 9}, 895 (1995).

\bibitem{dba98} D. ben-Avraham, ``Complete exact solution of
diffusion-limited coalescence, $A+A\rightarrow A$," {\it Phys. Rev. Lett.}
{\bf81}, 4756 (1998).

\bibitem{torney83} D. C. Torney and H. M. McConnell, ``Diffusion-limited
reactions in one dimension," {\it J. Phys. Chem.} {\bf87}, 1941 (1983).

\bibitem{balding88} D. Balding, ``Diffusion-reaction in one dimension,"
{\it J. Appl. Prob.} {\bf 25}, 733 (1988).

\bibitem{spouge88} J. L. Spouge, ``Exact solutions for a
diffusion-reaction process in one dimension," {\it Phys. Rev. Lett.} {\bf
60}, 871 (1988).

\bibitem{henkel} M. Henkel, E. Orlandini, and G. M. Sch\"utz,
``Equivalence between stochastic systems," {\it J. Phys. A}
{\bf28}, 6335 (1995);  M. Henkel, E. Orlandini, and J. Santos, 
``Reaction-diffusion processes from equivalent integrable quantum chains,"
{\it Ann. Phys. (NY)} {\bf259}, 163 (1997).

\bibitem{simon} H. Simon, ``Concentration for one and two 
species one-dimensional reaction-diffusion systems," {\it J. Phys. A}
{\bf28}, 6585 (1995).

\bibitem{droz} D. Balboni, P.-A. Rey, and M. Droz, ``Universality of a
class  of annihilation-coagulation models," {\it Phys. Rev. E} {\bf52},
6220 (1995).

\bibitem{bramson} M. Bramson and D. Griffeath, ``Clustering and
dispersion for interacting particles," {\it Ann. Prob.} {\bf 8}, 183
(1980);  {\it Z. Wahrsch. Geb.} {\bf 53}, 183 (1980).

\bibitem{peliti} L. Peliti, ``Path integral approach to birth-death
processes on a lattice," {\it J. Phys., Paris} {\bf 46}, 1469 (1985);
{\it J. Phys. A} {\bf 19}, L365 (1985).

\bibitem{krapivsky93} P.~L.~Krapivsky, ``Exact solutions for
aggregation-annihilation processes in one dimension," {\it Physica\ A}
{\bf 198}, 150 (1993); ``Aggregation-annihilation processes with
injection," {\it Physica\ A} {\bf 198}, 157 (1993).

\bibitem{privman} V. Privman, ``Exact results for diffusion-limited
reactions with synchronous dynamics," {\it Phys. Rev. E} {\bf 50}, 50
(1994); ``Exact  results for 1D conserved order parameter model," {\it
Mod. Phys. Lett. B} {\bf 8}, 143 (1994); V. Privman, A.M.R. Cadilhe, and
M.L. Glasser, ``Exact  solutions of anisotropic diffusion-limited
reactions with coagulation and annihilation," {\it J. Stat. Phys.} {\bf
81}, 881 (1995); ``Anisotropic diffusion-limited reactions  with
coagulation and annihilation," {\it Phys. Rev. E} {\bf 53}, 739 (1996).

\bibitem{glaser} For a review of field-theoretical techniques, and
additional references, see: D.~C.~Mattis and M.~L.~Glasser, ``The uses of
quantum field-theory in diffusion-limited reactions," {\it Rev. Mod.
Phys.} {\bf 70}, 979 (1998).

\bibitem{racz85} Z. R\'acz, ``Diffusion-controlled annihilation in the
presence of particle sources: exact results in one dimension," {\it Phys.
Rev. Lett.} {\bf 55}, 1707 (1985).

\bibitem{lushnikov87} A. A. Lushnikov, ``Binary reaction $1+1\to
0$ in one dimension," {\it Phys. Lett. A} {\bf120}, 135 (1987).

\bibitem{martin95} H. O. M\'artin, J. L. Iguain and M. Hoyuelos, ``Steady
state of imperfect annihilation and coagulation reactions," {\it J. Phys.
A} {\bf28}, 5227 (1995).

\bibitem{rey97} P. Rey and M. Droz, ``A renormalization group study of a
class of reaction-diffusion models, with particles input," {\it J. Phys.
A} {\bf30}, 1101 (1997).

\bibitem{grassberger84} P. Grassberger, F. Krausse, and T. von der Twer,
``A new type
of kinetic critical phenomenon," {\it J. Phys. A} {\bf17}, L105 (1984).

\bibitem{grassberger89} P. Grassberger,  ``Some further results on a
kinetic critical phenomenon," {\it J. Phys. A} {\bf22}, L1103 (1989).

\bibitem{sudbury90} A. Sudbury, ``The branching annihilating process: An
interacting particle system," {\it Ann. Prob.} {\bf18}, 581 (1990).

\bibitem{takayasu92} H. Takayasu and A. Y. Tretyakov, ``Extinction,
survival, and dynamical phase transition of branching annihilating random
walkers," {\it Phys. Rev. Lett.} {\bf68}, 3060 (1992).

\bibitem{jensen93} I. Jensen, ``Conservation laws and universality in
branching annihilating random walks," {\it J. Phys. A} {\bf26}, 3921
(1993).

\bibitem{dba94} D. ben-Avraham, F. Leyvraz, and S. Redner, ``Propagation
and extinction in branching annihilating random walks,"
{\it Phys. Rev. E} {\bf 50}, 1843 (1994).

\bibitem{mussawisade98} K. Mussawisade, J. E. Santos, and G. M. Sch\"utz,
``Branching annihilating random walks in one dimension: some exact
results," {\it J. Phys. A} {\bf31}, 4381 (1998). 

\bibitem{szabo99} G. Szabo and M. A. Santos, ``Branching annihilating
random walks with parity conservation on a square lattice,"
{\it Phys. Rev. E} {\bf 59}, R2509 (1999).

\bibitem{alemany95} P. A. Alemany and D. ben-Avraham, ``Inter-particle
distribution functions for one-species diffusion-limited annihilation,
$A+A\rightarrow 0$", {\it Phys. Lett. A}  {\bf206}, 18 (1995).

\bibitem{dba00} D. ben-Avraham and S. Havlin, {\it Diffusion and
Reactions in Fractals and Disordered Systems}, (Cambridge University
Press, 2000).

\bibitem{masser00} T. Masser and D. ben-Avraham, ``Kinetics of
coalescence, annihilation, and the $q$-state Potts model in one
dimension," {\it Phys. Lett. A} {\bf275}, 382 (2000).

\bibitem{henkel01} For a novel application to $A+A\to A$ and $2A\to3A$
($A0A\to  AAA$), see: M. Henkel and H. Hinrichsen, ``Exact solution of a
reaction-diffusion process with three-site interactions," e-preprint:
cond-mat/0010062, to be published in {\it J. Phys. A} (2001).

\bibitem{habib} After submission of this manuscript, we learned of
beautiful, similar work, done independently: ``Diffusion-limited reaction
in one dimension: paired and unpaired nucleation," S.~Habib,
K.~Lindenberg, G.~Lythe, and C.~Molina-Par{\'\i}s, {\it preprint}.

\bibitem{abad99} E. Abad, H. L. Frisch, and G. Nicolis, ``1D lattice
dynamics of the diffusion limited reaction $A+A\rightarrow A+S$: transient
behavior," {\it J. Stat. Phys.} {\bf99}, 1397 (2000).

\bibitem{AS} M.~Abramowitz and I.~M.~Stegun, eds., {\it Handbook of
Mathematical Functions},  (Dover Publications, New York, 1964).

\bibitem{burschka} M.~A.~Burschka, C.~R.~Doering, and D.~ben-Avraham,
``Transition in the relaxation dynamics of a reversible  diffusion-limited
reaction," {\it Phys. Rev. Lett.} {\bf 63}, 700 (1989); C.~R.~Doering and
M.~A.~Burschka, ``Long crossover time in a finite system," {\it Phys. Rev.
Lett.} {\bf 64}, 245 (1990).

\bibitem{dba} D. ben-Avraham, ``Fisher waves in the diffusion-limited
coalescence process $A+A\rightleftharpoons A$," {\it Phys. Lett. A}
{\bf 247}, 53 (1998); ``Diffusion-limited
coalescence, $A+A\rightleftharpoons A$, with a trap," {\it Phys. Rev. E}
{\bf 58}, 4351 (1998); ``Inhomogeneous steady-states of
diffusion-limited coalescence, $A+A\to A$," {\it Phys. Lett. A}
{\bf 249}, 415 (1998); A. Donev, J. Rockwell, and
D.~ben-Avraham, ``Generalized von Smoluchowski model of reaction rates, 
with reacting particles and a mobile trap,"
{\it J. Stat. Phys.} {\bf 95}, 97 (1999).


\end{thebibliography}
\end{document}